\documentclass{aa}
\begin{document}
%\thesaurus{05 ()}
\title{Simultaneous time-series spectroscopy and multi-band photometry of the 
sdBV PG\,1605$+$072
\thanks{Based on observations obtained at the German-Spanish Astronomical 
Center (DSAZ), Calar Alto, operated by the Max-Planck-Institut f\"ur Astronomie 
Heidelberg jointly with the Spanish National Commission for Astronomy.}}
\author{S. Falter\inst{1} \and U. Heber\inst{1} \and S. Dreizler\inst{2} \and 
S. L. Schuh\inst{2} \and O. Cordes\inst{3} \and H. Edelmann\inst{1}} 
\offprints{S. Falter,\\
email: \tt{falter@sternwarte.uni-erlangen.de}}
\institute {Dr. Remeis-Sternwarte, Astronomisches Institut der Universit\"at
Erlangen-N\"urnberg, Sternwartstr. 7,\\D-96049 Bamberg, Germany
\and
Institut f\"ur Astronomie und Astrophysik, Universit\"at T\"ubingen, Sand 1, 
D-72076 T\"ubingen, Germany
\and
Sternwarte der Universit\"at Bonn, Auf dem H\"ugel 71, D-53121 Bonn, Germany}
%\date{Received,  accepted }
\date{}
\abstract{
We present time-series spectroscopy and multi-band photometry of the sdBV 
PG\,1605$+$072 carried out simultaneously at the Calar Alto 2.2m and 3.5m
telescopes. The periodogram analysis of the radial velocity curves reveals three
frequencies at 2.078, 2.756, and 1.985 mHz for $\mathrm{H}_{\beta}$ and at 2.076,
2.753, and 1.978 mHz for $\mathrm{H}_{\gamma}$. The corresponding radial velocity
amplitudes are 12.7, 8.0, and 7.9 km/s for $\mathrm{H}_{\beta}$ and 14.3,
6.5, and 7.2 km/s for $\mathrm{H}_{\gamma}$. Furthermore, we found five
frequencies that are present in all wavelength bands of the BUSCA photometer.
The frequencies detected in the radial velocity curves are recovered by the 
photometric measurements. Moreover, additional frequencies were present in the
periodograms which could not be identified in all four bands simultaneously. The
comparison of the amplitudes presented here with previous results from radial
velocity and photometric observations of PG\,1605$+$072 shows a significant
change or even switch in the power of the modes within short time scales, i. e.
about one year. No changes in frequency were registered and the phases of 
the modes show no wavelength dependency within our multi-band photometry.
}

\maketitle

\section{Introduction\label{intro}} 
 
Subluminous B (sdB) stars dominate the populations of faint blue 
stars of our own Galaxy and are found in the disk (field sdBs) as well as in 
globular clusters (Moehler et al. \cite{mohe97}). Furthermore, these stars play 
an important role trying to explain the ``UV upturn phenomenon'' observed in 
elliptical galaxies and galaxy bulges (Greggio \& Renzini \cite{grre90}, 
\cite{grre99}). According to observations with the Ultraviolet Imaging Telescope 
(Brown et al. \cite{brfe97}) and the Hubble Space Telescope (Brown et al. 
\cite{brbo00}) sdB stars are sufficiently numerous to be responsible for the 
excessive UV flux.

It is generally accepted that sdB stars can be identified with models of the 
extended Horizontal Branch (EHB) 
burning He in their core (Heber \cite{hebe86}; Saffer et al. \cite{sabe94}). 
The hydrogen envelope surrounding the core of about half a solar mass is very 
thin ($<$\,2\,\% by mass) and therefore inert. These EHB stars will continue their  
evolution directly towards the white dwarf graveyard avoiding the AGB and 
planetary nebula phases (Dorman et al. \cite{doro93}). 

How sdB stars evolve towards the EHB with effective temperatures of up to 
40\,000\,K remains a puzzle. The star must have lost all but a tiny fraction of 
the hydrogen envelope at the same time as the He core has attained the minimum 
mass ($\approx$\,0.5\,M$_\odot$) required for the He flash. This challenges every 
mass loss mechanism in aspects of timing and effectivity. Recent findings 
(Maxted et al. \cite{mahe01}; Saffer et al. \cite{sagr01}; Heber et al. 
\cite{hemo02}) emphasize the significance of close binary evolution. 

Some of the sdB stars were recently found to exhibit rapid multi-periodic light
variations (P $\approx$ 80\,--\,600 s) of low amplitudes (a few mmag). They form 
a new class of pulsating stars named after the prototype EC\,14026 stars
\footnote{EC\,14026 stars are now officially V361 Hya stars} (Kilkenny et 
al. \cite{kilk97}). Since then a relatively large number of new sdB pulsators 
has been discovered. 31 are known today (Charpinet \cite{char01}, Piccioni et 
al. \cite{picc00}, Silvotti et al. \cite{silv00}). The 
observed brightness variations are caused by radial and non-radial, low degree 
and low order acoustic pulsation modes. The pulsations in these stars are driven 
by an opacity bump due to Fe and other metallic species (Charpinet et al. 
\cite{chfo97}) at a temperature of $\approx$ 2\,$\times$\,$10^5$\,K in the sdB 
envelope.

Stellar pulsations allow a direct insight into the structure of such stars and 
therefore into the evolutionary history. The frequencies or periods of the 
pulsation modes probe the chemical stratification and the mass which 
otherwise are difficult or even impossible to determine.
The power of asteroseismological tools has been demonstrated in the field of 
pulsating white dwarfs for which stellar parameters like mass, luminosity or 
thickness of the envelope were derived (e.~g. Winget et al. \cite{wing91}). In 
the case of variable sdB stars these parameters will constrain the evolutionary 
history and consequently shed more light on the origin of these stars.

Identification of pulsation modes (characterized by spherical harmonics with 
the indices $l$ and $m$) is a prerequisite for asteroseismology. Brassard et el.
(\cite{bras01}) have successfully carried out an asteroseismological analysis
for PG\,0014$+$067. For the first time they were able to determine the stellar
mass ($\rm M_{\star}/M_{\odot} = 0.490\pm 0.019$) as well as the envelope mass
($\rm log\left(M_{env}/M_{\odot}\right) = -4.31 \pm 0.22$) and both are in 
excellent agreement with predictions from evolutionary models (Dorman et al. 
\cite{doro93}). 

Pulsations produce not only photometric variations but also line profile 
variations that offer an alternative approach towards mode identification. 
PG\,1605$+$072 is the ideal target for this application: it has the longest 
pulsation periods ($\approx$\,500 s) which enables spectra with reasonable S/N to be
obtained within each pulsation period. Moreover, this star has the largest 
variations of all known sdBVs (0.2\,mag in the optical) and by far the richest 
frequency spectrum ($>$ 50 modes, Kilkenny et al. \cite{kilk99}). A recent 
spectroscopic study revealed this star to be a rather rapid rotator 
(v\,$\sin i = $ 39 km/s, Heber et al. \cite{here99}) which will complicate the 
identification of modes due to non-linear effects on mode splitting.

This work presented here serves as a feasibility study in order to find out
whether an asteroseismologic analysis of PG\,1605$+$072 is possible. For this
purpose we have done simultaneously time-series spectroscopy and multi-band
photometry (observations and reductions in Sect.~\ref{obsred}). The analysis of 
the results is presented in Sect.~\ref{anal}. Previously, other groups have done
radial velocity studies (O'Toole et al. \cite{otoo00}, \cite{otoo02}, Woolf et
al. \cite{wool02}) or photometric campaigns (e.~g. Kilkenny et al. \cite{kilk99}) 
on PG\,1605$+$072. Simultaneous multi-band photometry has not been observed 
before. This enables us to study the temporal evolution of the frequencies and
amplitudes of the pulsation modes (Sect.~\ref{comp}). Finally, we discuss our 
results and give a brief outlook to future work in Sect.~\ref{conc}.

\section{Observations and data reduction\label{obsred}}

\subsection{Time-series spectroscopy}

We obtained time resolved longslit spectra for the sdBV PG\,1605$+$072 during
one observing run. The observations were carried out with the 3.5m telescope at 
the DSAZ, Calar Alto, Spain and the TWIN spectrograph (two SITe-CCDs with 
2048\,$\times$\,800 pixels \`a 15 $\mu$m). The time allocation committee awarded two 
nights on May 14 and 
15, 2001. Because of rain during the first night data were only obtained during 
the second night (5h\,27\,m of data). The passage of clouds and variable seeing 
affected the quality of the data considerably. In order to achieve an adequate 
time resolution observations were carried out in trailing mode which means that 
the telescope tracking is tuned such that the star moves slowly along the slit in 
N-S direction. The drift velocity was chosen at 270\arcsec\ per hour so that a 
time resolution of 15 s per pixel resulted. The time resolution is dependent on 
the seeing conditions and thus varied during the night. The slit width was 
1.5\arcsec . We chose the gratings T05 and T06 (36 $\mathrm{\AA}$/mm) 
corresponding to a spectral resolution of $\approx$ 1 $\mathrm{\AA}$. The spectral 
ranges from 3860 -- 4960 $\mathrm{\AA}$ in the blue and from 5880 -- 6920 
$\mathrm{\AA}$ in the red were covered. Moreover, to decrease the readout time 
and the noise level the data were binned (1$\times$2).

\subsection{Multi-band photometry}

\begin{figure}
\vspace{11cm}
\includegraphics{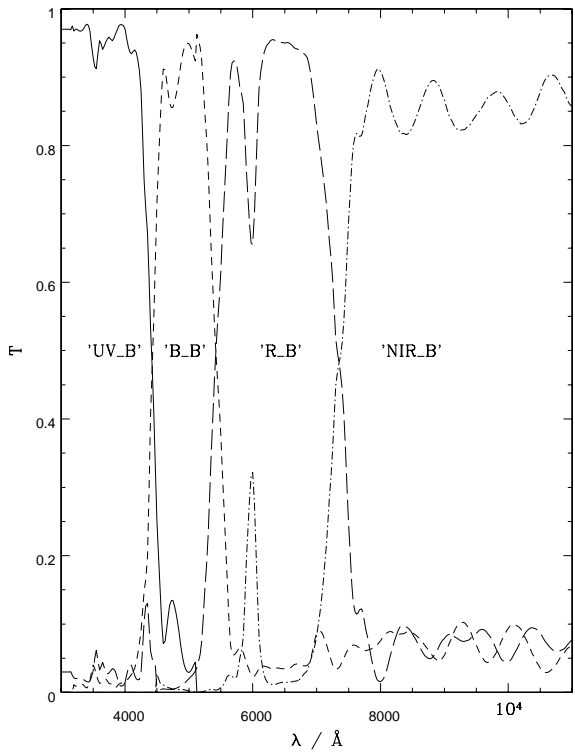}
\caption{Transmission curves of the four BUSCA wavelength bands. The four ranges
are denoted by '$UV_{\mbox{\footnotesize B}}$' (central wavelength: 3\,600 $\mathrm{\AA}$), '$B_{\mbox{\footnotesize B}}$' (4\,800 
$\mathrm{\AA}$), '$R_{\mbox{\footnotesize B}}$' (6\,300 $\mathrm{\AA}$) and '$NIR_{\mbox{\footnotesize B}}$' (8\,000 $\mathrm{\AA}$) 
although these are not comparable to the classical Johnson UBVRI system.}
\label{beam}
\end{figure}

Simultaneously with the time resolved spectroscopy of the target star its
light curve was measured in four wavelength bands at the 2.2m telescope at Calar 
Alto. At this telescope five nights between May 14 and 18, 2001 were awarded.
No data could be taken during the first night and from the middle of the fourth
night until the end of the run due to bad weather conditions. 12\,h\,27\,m of 
pure 
observation time resulted in 880 measurements. The telescope was equipped with 
the new multi-band photometer BUSCA (Reif et al. \cite{reif99}, Cordes et al.
in prep.) which is able to measure in four photometric bands 
simultaneously. Three beamsplitters split the incoming light into four beams 
that feed four 4k$\times$4k CCDs. Only Str{\o}mgren filters were available for 
BUSCA. 
Use of these filters would have further reduced the S/N. Therefore, no filters 
were inserted and the beamsplitters served as non-overlapping broad band 
filters. The transmission curves of the four bands are shown in Fig.~\ref{beam}.
These curves convoluted with the quantum efficiency curves of the CCDs 
provide the pass bands for our observations. The four wavelength bands are
denoted by '$UV_{\mbox{\footnotesize B}}$', '$B_{\mbox{\footnotesize B}}$', '$R_{\mbox{\footnotesize B}}$' and '$NIR_{\mbox{\footnotesize B}}$' throughout the paper following 
the notation of Cordes et al. (in prep.). We chose an exposure time of 15\,s
which provides a good S/N and a reasonable coverage of the main frequencies.
%The relatively short pulsation 
%periods require short exposures (15 s) to cover one period with as many data 
%points as possible. 
2$\times$2 binning (resolution: 0.17\arcsec\ per unbinned 
pixel) and the windowing option (reading out 346$\times$400 pixels of the CCDs) 
helped to reduce the full cycle time to 51 s.
%Reading out the CCDs without 
%windowing would take 2\,m\,20\,s (unbinned: $\approx$ 8\,m).

The brightness variations of PG\,1605$+$072 were detected by relative
photometry. For this purpose two stars of similar brightnesses in the R band 
($\mathrm{R}_{1}$\,=\,13.2 and $\mathrm{R}_{2}$\,=\,13.4) next to our target star 
were used.

\subsection{Data reduction}

Data reduction was done using the IDL based software SPEX ({\bf L}ong slit {\bf 
SP}ectrum {\bf EX}traction package\footnote{see http://astro.uni-tuebingen.de/$\sim$schuh/spex/index.html}), 
and TRIPP ({\bf T}ime {\bf R}esolved {\bf I}maging {\bf P}hotometry 
{\bf P}ackage, see Schuh et al. \cite{schu99}). TRIPP is based on the CCD 
photometry routines written by R.D. Geckeler (\cite{geck98}) and performs 
aperture photometry. The most important 
step is the determination of the relative flux of the target star with one or 
more stars as comparison objects. The detection of variations of the order of a 
few mmag is only possible when comparison stars and sky background are recorded 
simultaneously (especially under comparatively poor conditions).

SPEX is designed for the reduction of long slit spectra. The reduction comprises 
of the standard procedures flatfielding, biasing, cosmic ray event extraction,
and 2-D wavelength calibration. The latter is very important because we extract
spectra from single CCD rows in order to establish the time-series. Therefore, 
we have to ensure that the position of the comparison lines perpendicular to 
the dispersion direction doesn't change. In fact, these positions vary and thus 
we correct for that by fitting a polynomial. After the calibration of one CCD 
row the resulting dispersion relation is applied to every single CCD row.

One crucial point was the calibration of the time axis. As a consequence of the 
trailing mode observation the seeing disk moves slowly  
across the CCD and therefore the signal smears over the trailing direction. The
initial point of time of the measurement and the cycle time were determined by
fitting Gaussian profiles to the rising flank of the star's signal. Finally, 
radial velocities were derived by fitting Lorentz profiles to the Balmer lines
$\mathrm{H}_{\alpha}$-$\mathrm{H}_{8}$ as well as to \ion{He}{i} 4471 
$\mathrm{\AA}$ and \ion{He}{ii} 4686 $\mathrm{\AA}$.

\section{Analysis of the time-series\label{anal}}

Both the spectroscopic as well as the photometric time-series were analysed with
the program package TRIPP. It enables the calculation of periodograms,
confidence levels and fits with multiple sine functions (see Dreizler et al.
\cite{drei02} for more details).

%The single site data of time-series that span several days or even weeks 
%problems. TRIPP is able to handle these unevenly sampled data by using a method
%that was introduced by Scargle (\cite{scar82}). We calculate the so called 
%Lomb-Scargle periodograms from the radial velocity curves and light curves.
%Scargle's definition provides a linear least-square fit of the form 
%$A\cos\omega t + B\sin\omega t$ for all frequencies $\omega$ of
%interest. Press \& Rybicki (\cite{prry89}) have made available a fast
%implementation that keeps computation time modest for large data sets. Using
%this fast algorithm also allows to calculate a statistical estimate of detection
%limits for every single run. For this purpose a large number of light curves with
%similar timing as during the observations are simulated. Additionally, we assume
%white noise of the measured variance of the given observation. The highest
%peak of every individual periodogram is stored and after that all these peaks are
%monotonically sorted. The height of the simulated peaks serves as a reference
%for the height of the peaks from the observation's periodogram. The confidence
%levels are derived in the following way: In the case when an observed peak at 
%any arbitrary frequency point is stronger than 90 \% of the simulated peaks
%(collected at the whole frequency range), the probability for it being a true
%detection is larger than 90 \%.

\subsection{Spectroscopy}

\begin{figure}
\vspace{11cm}
\includegraphics{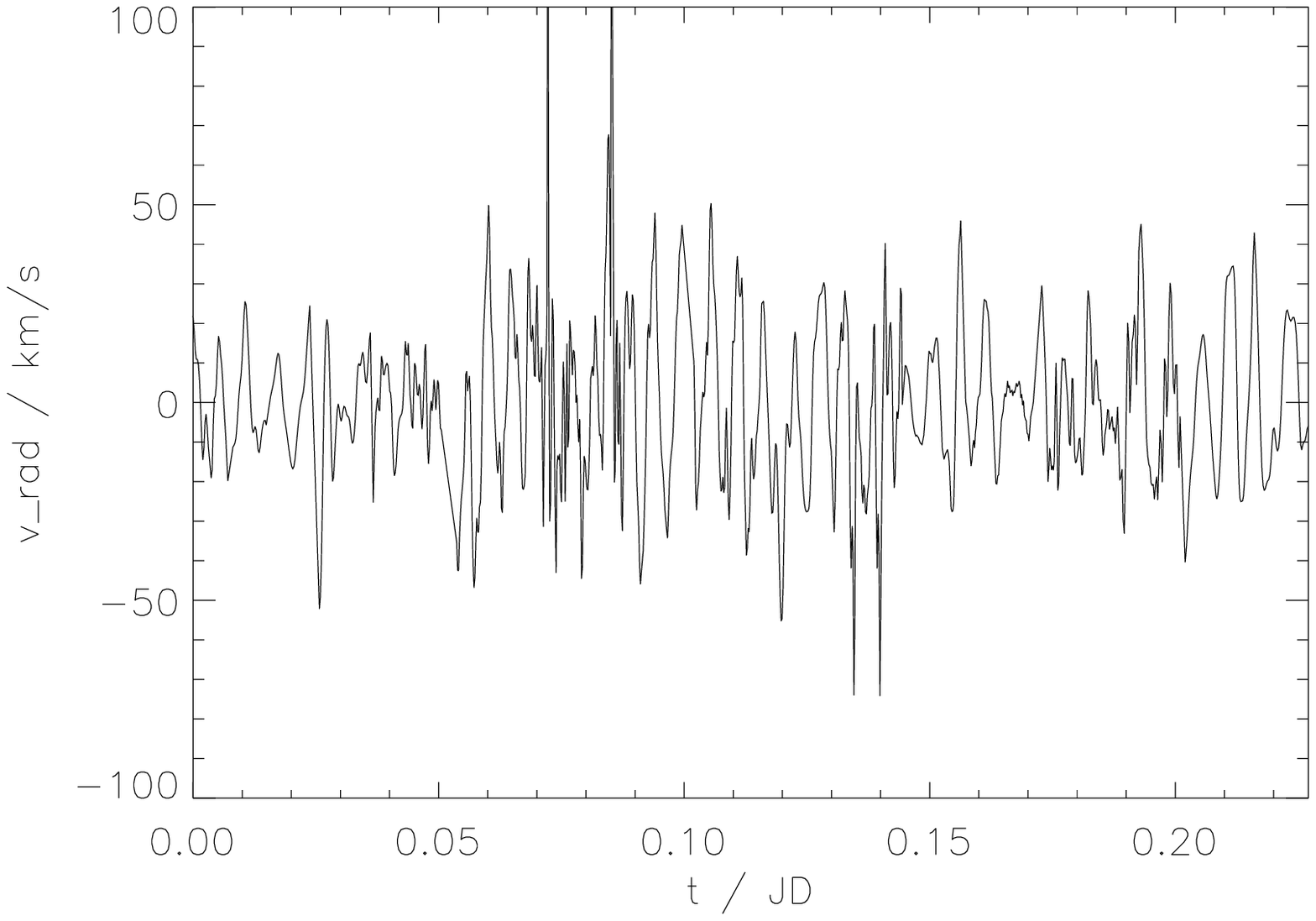}
\includegraphics{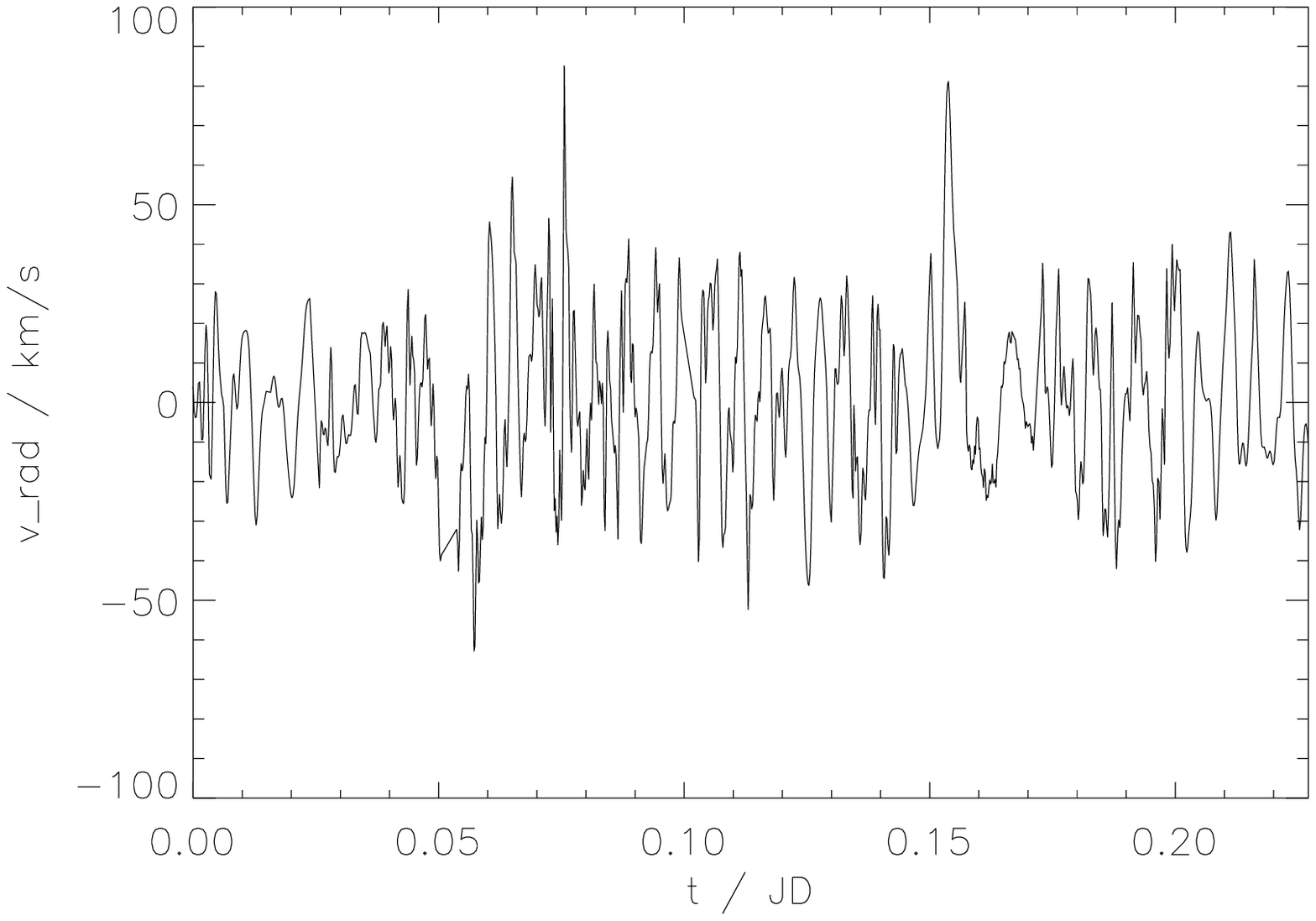}
\caption{Radial velocity curve for $\mathrm{H}_{\beta}$ (upper panel) and 
$\mathrm{H}_{\gamma}$ (lower panel); artefacts due to poor weather conditions 
were not removed}
\label{RVHb}
\end{figure}

We were able to derive accurate radial velocity curves for the Balmer lines 
$\mathrm{H}_{\beta}$ and $\mathrm{H}_{\gamma}$ (see Fig.~\ref{RVHb}). The 
frequency resolution for this run is calculated to be 51 
$\mu$Hz. Trailing mode observations require stable weather conditions so that 
the intensity of the star's signal is almost constant during one exposure. Thus 
bad seeing and transparency changes due to the passage of small clouds influence 
the quality of the data rather strongly. Because of these disturbances the 
quality of the radial velocity curves varies with time. The radial velocity
curves of the other Balmer lines and the He lines turned out to be too noisy for
a quantitative analysis and are not further discussed. Better conditions should
allow a more extensive study of radial velocity changes in this star.

\begin{figure}
\vspace{11cm}
\includegraphics{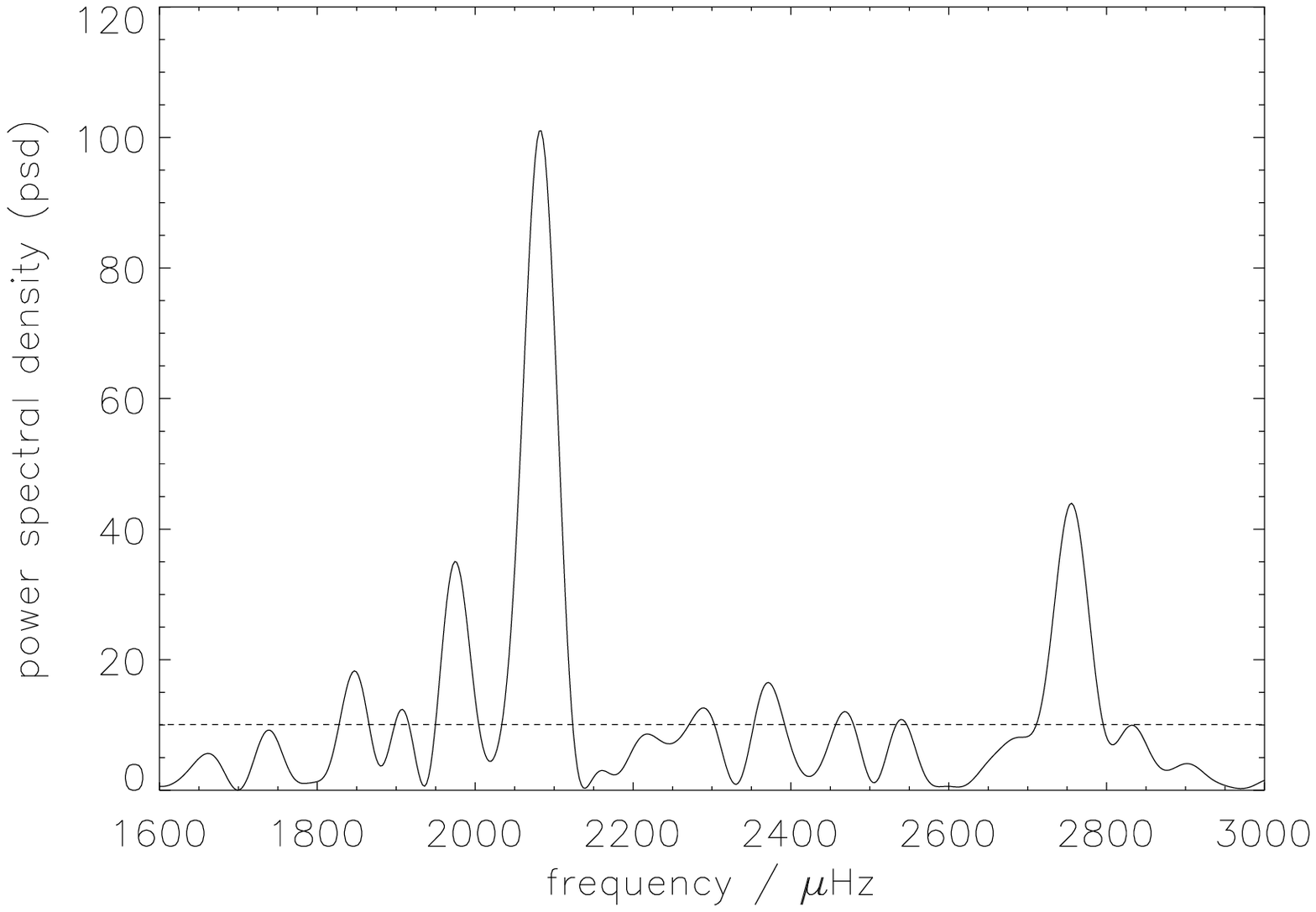}
\includegraphics{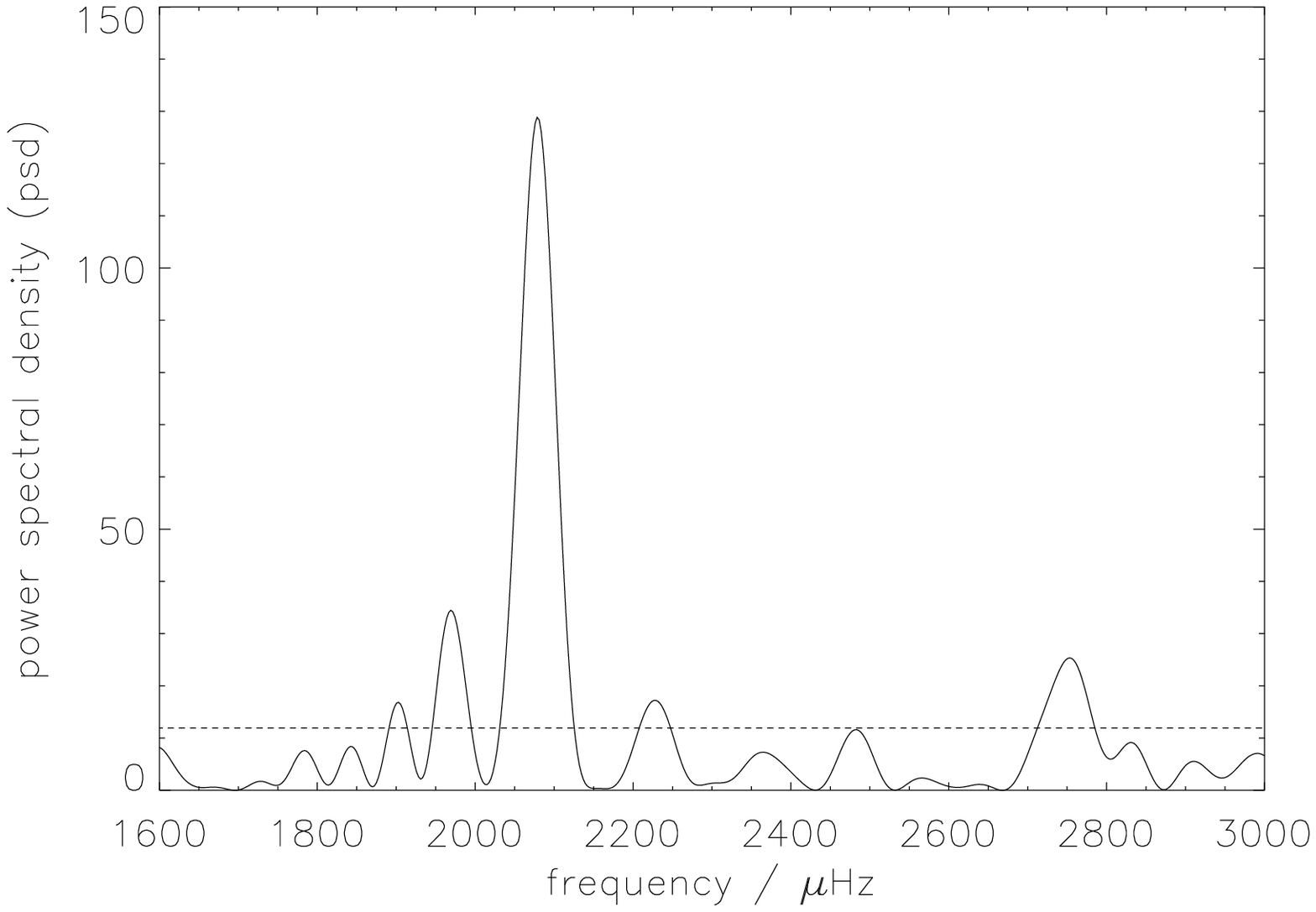}
\caption{Lomb-Scargle periodogram of the radial velocity curve of 
$\mathrm{H}_{\beta}$ (upper panel) and $\mathrm{H}_{\gamma}$ (lower panel). The 
power spectral density (psd) is a measure for the probability that a period is 
present in the radial velocity curve or light curve. The horizontal line 
represents the confidence level of 99 \% (3$\sigma$-level).}
\label{LSpHb}
\end{figure}

The Lomb-Scargle periodograms calculated from the radial velocity curves (see
Fig.~\ref{LSpHb}) show peaks clearly surpassing the 3$\sigma$ confidence
level. Applying a prewhitening procedure to the data three frequencies and their
amplitudes (see Table \ref{radvel}) were found for both Balmer lines 
investigated. Other peaks that seem
to lie above the 3$\sigma$ level could not be extracted by our procedure. The
dominant frequency at 2.076 mHz has the largest velocity amplitude with 12.7 
km/s for $\mathrm{H}_{\beta}$ and 14.3 km/s for $\mathrm{H}_{\gamma}$. The 
amplitudes of the other two frequencies are lower than the strongest one with 
values ranging from 6 to 8 km/s. We determined the amplitude accuracy 
$\Delta$A by calculating the 
median value of the white noise in the frequency range 3\,--\,7 mHz where almost 
no power arises in the frequency spectrum. According to this, we found
$\Delta$A\,=\,1.5\,km/s for $\mathrm{H}_{\beta}$ and $\Delta$A\,=\,1.0\,km/s for 
$\mathrm{H}_{\gamma}$.

\subsection{Multi-band photometry}

BUSCA is a unique instrument which enables the measurement in four different
wave bands simultaneously. As a result we obtained four light curves.

\begin{figure}
\vspace{20cm}
\includegraphics{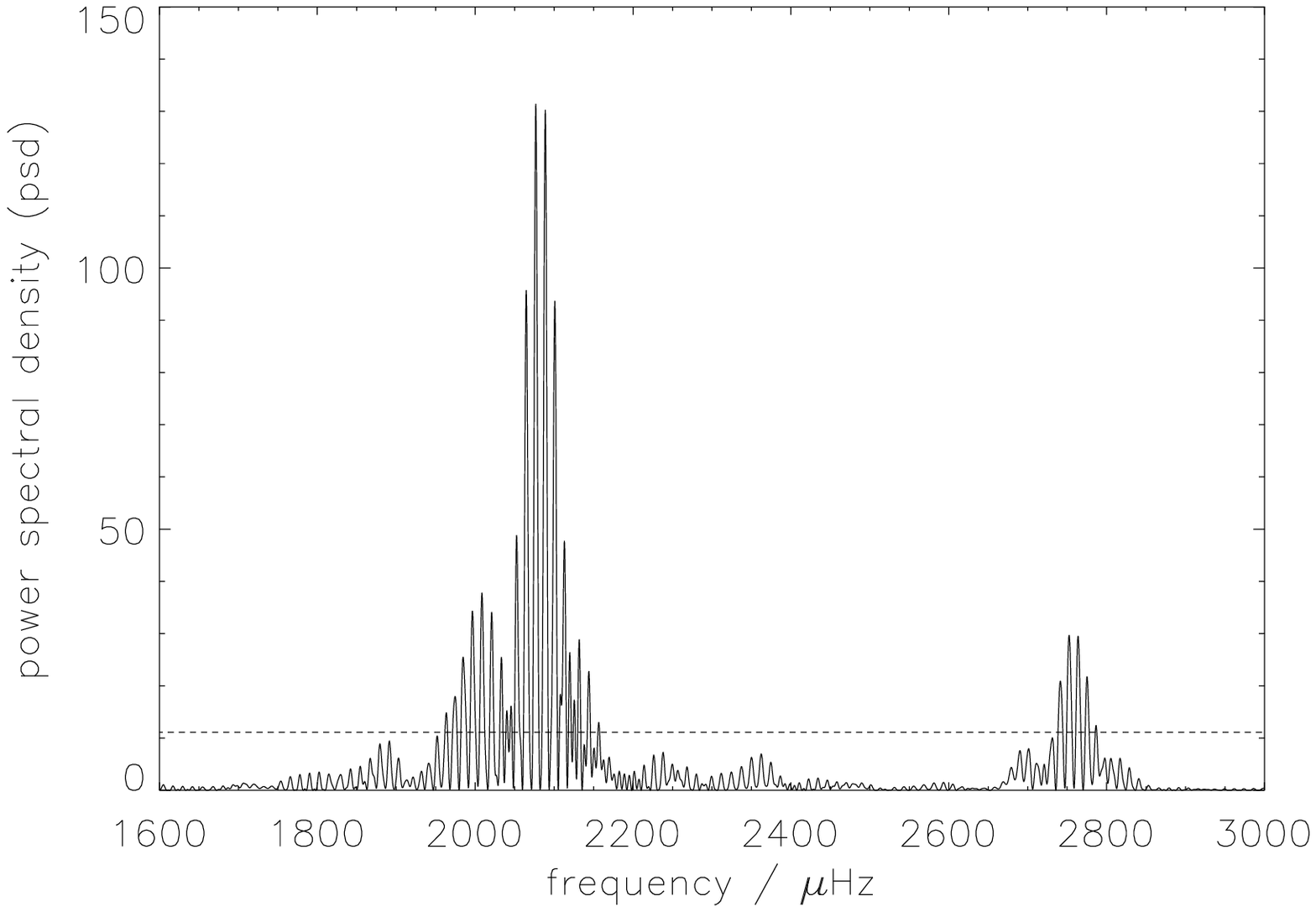}
\includegraphics{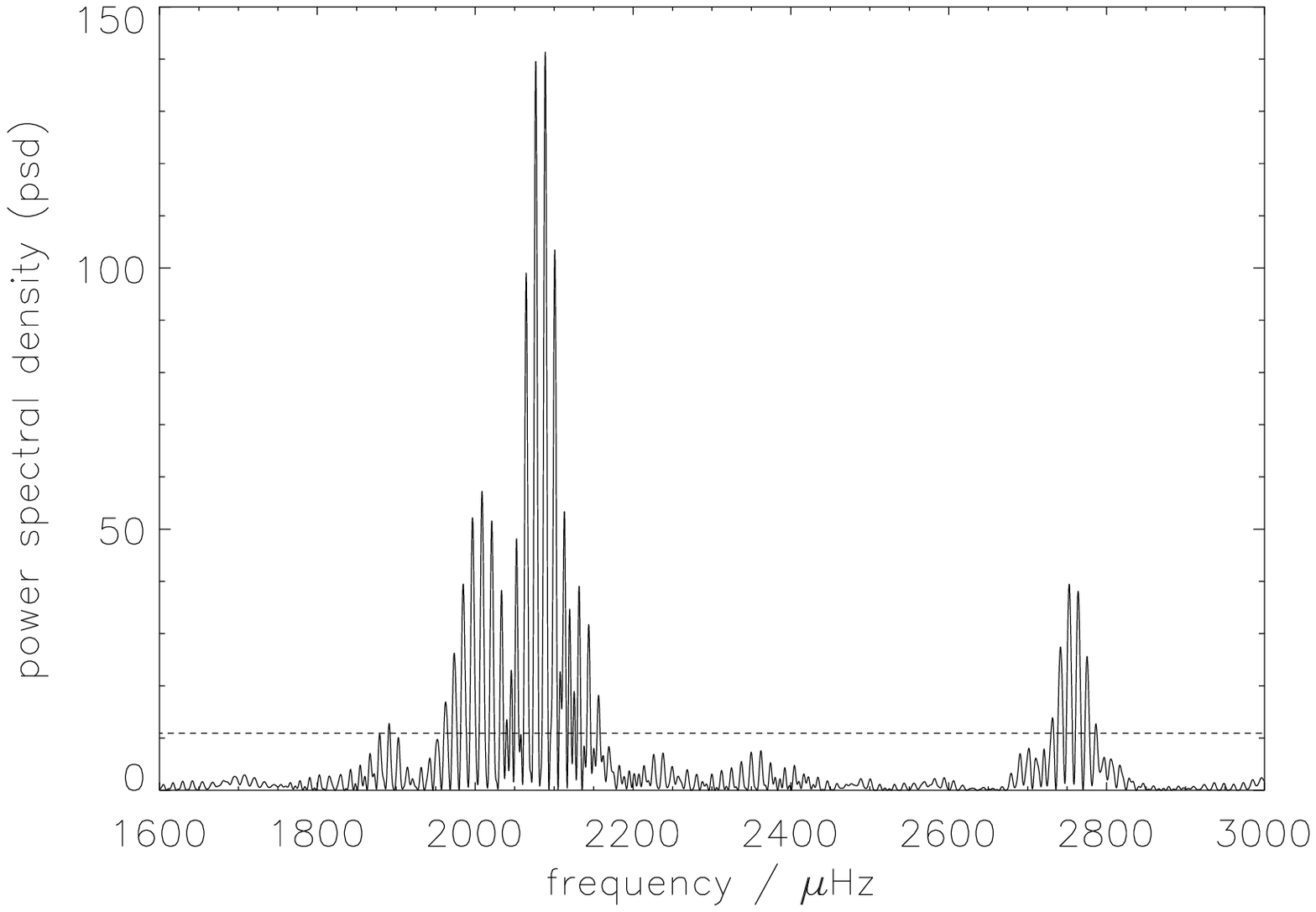}
\includegraphics{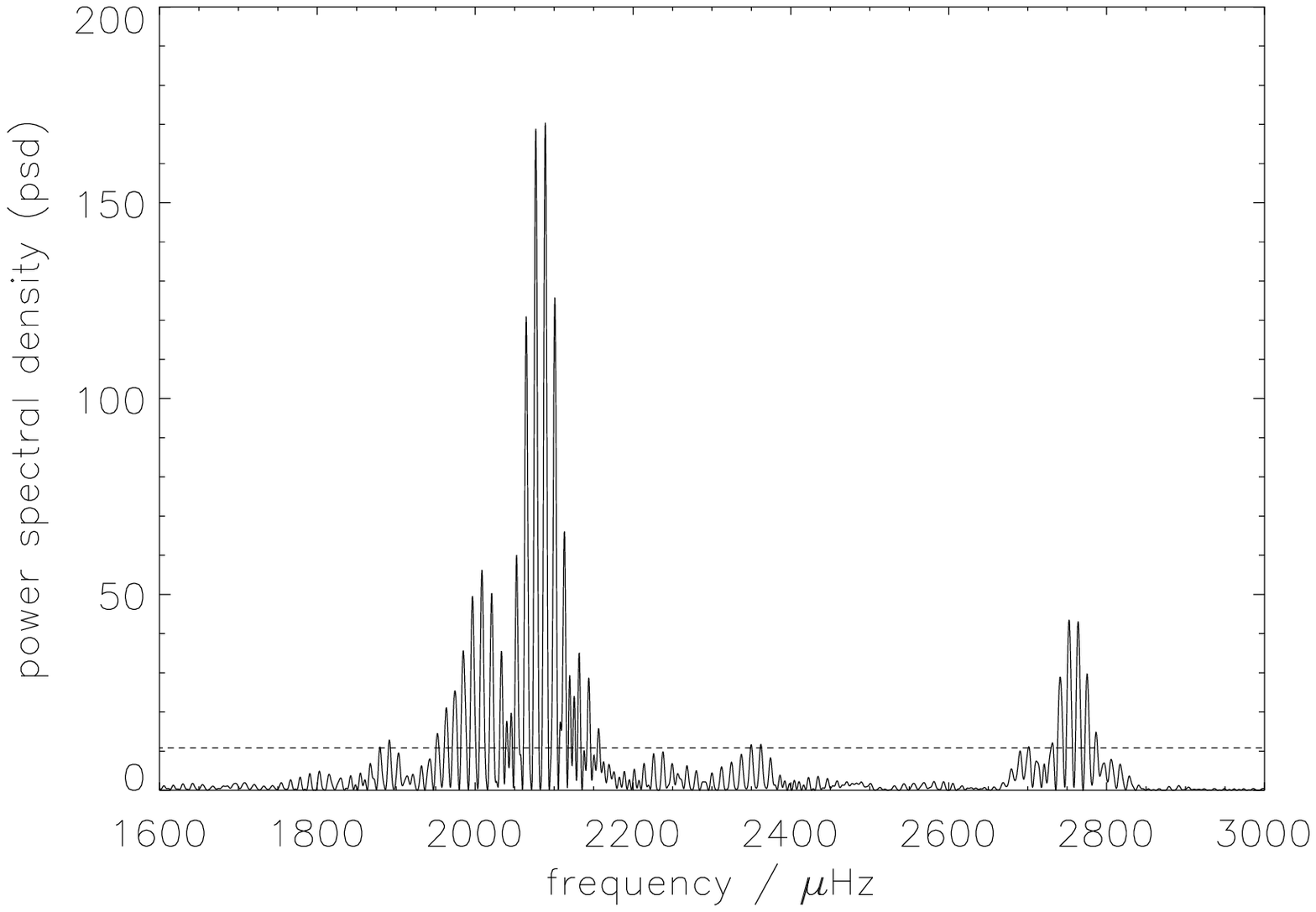}
\includegraphics{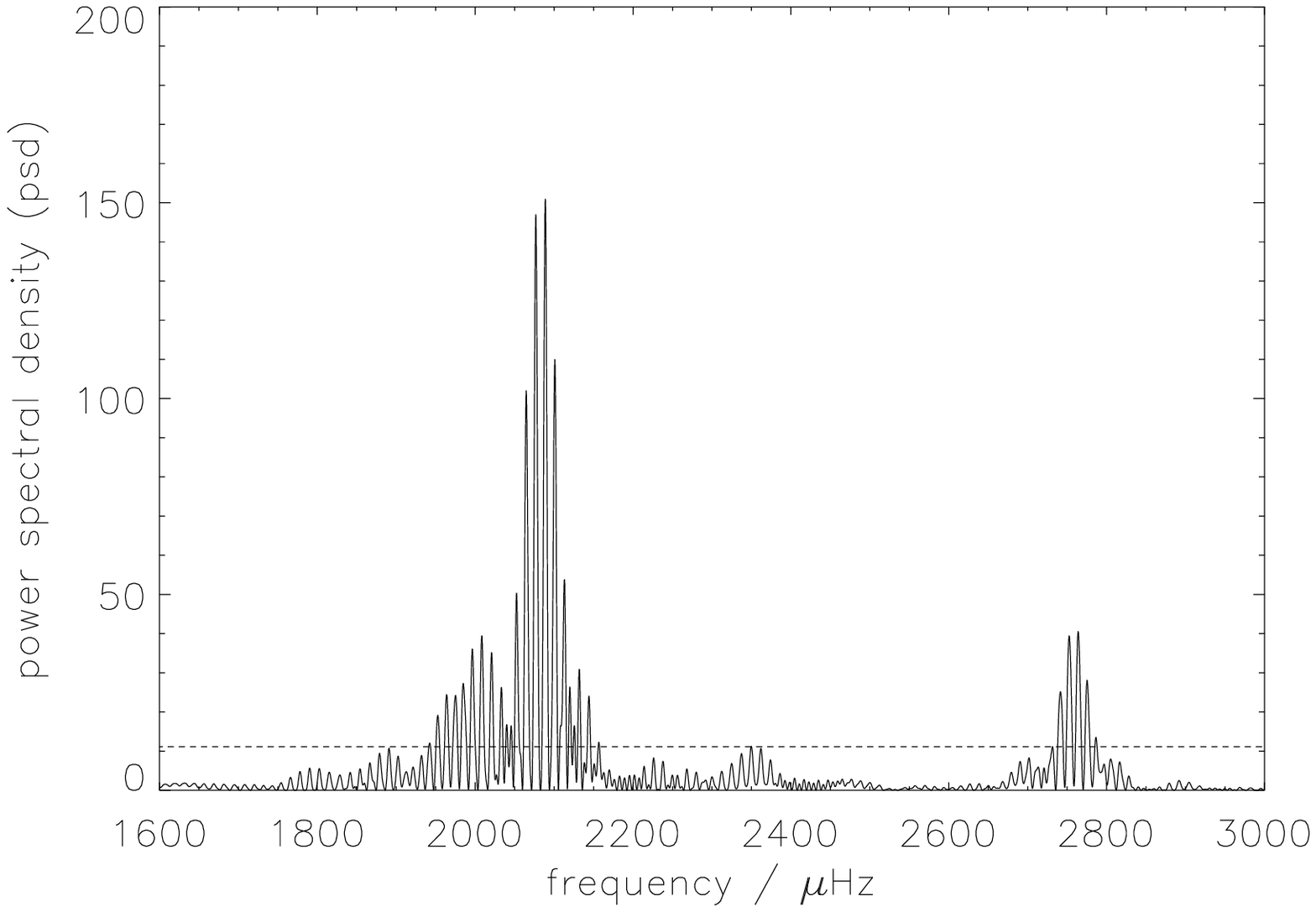}
\caption{Lomb-Scargle periodogram of the light curves of the '$UV_{\mbox{\footnotesize B}}$', '$B_{\mbox{\footnotesize B}}$', 
'$R_{\mbox{\footnotesize B}}$' and '$NIR_{\mbox{\footnotesize B}}$' band (from top to bottom) of the BUSCA camera. The 
horizontal line in each panel represents the confidence level of 99 \% 
(3$\sigma$-level).}
\label{LSpUV}
\end{figure}

Fig.~\ref{LSpUV} shows the Lomb-Scargle periodograms of all BUSCA bands. They 
are quite similar to the periodograms derived from the radial velocity curves
(Fig.~\ref{LSpHb}). This data set spans over three nights so that daily aliases 
are clearly visible and the frequency resolution is much better at $\Delta\nu$ 
= 5.68 $\mu$Hz (compared to 51 $\mu$Hz for the spectroscopy). Again we applied 
the prewhitening technique in order to remove all significant peaks from the 
periodogram and to obtain the amplitudes for each frequency. As 
before, the horizontal line in the diagrams represents the 3$\sigma$ 
confidence level above which we assume the detected frequencies to be real. In 
all four wave bands five peaks with the same frequencies can be identified 
(see Table \ref{freqb} and \ref{freqc}). The dominant frequency is again found 
at 2.076 mHz and therefore confirms the results from spectroscopy. Furthermore, 
additional frequencies were found in the region around 2.74\,--\,2.78 mHz but 
these peaks are closely spaced so that a corresponding identification in all 
BUSCA bands due to the medium frequency resolution was not possible. Peaks 
that fall below the 3$\sigma$ confidence level were not removed from the 
periodograms.

%The amplitudes of the brightness variations are measured in {\it modulation
%intensity} (= mi). The light curves are normalized by their mean value, i. e. the 
%unit mi is equivalent to the fraction of intensity 
%$\delta\mathrm{I}/\mathrm{I}$. It was first introduced by Winget et al. 
%(\cite{wing94}) and gives a linear measure of fractional intensity. The mi can 
%be directly transformed into magnitudes (1 mmi = 1.086 mmag), the latter will 
%be used throughout the whole paper.
{\bf The amplitudes of the brightness variations are measured in {\it fractional 
intensity}. The light curves are normalized to the fraction of intensity 
$\delta\mathrm{I}/\mathrm{I}$ and the amplitudes were then converted to mmag. 
This unit will be used throughout the whole paper. 
}
The accuracy of these amplitudes are
calculated in the same way as we did it for the radial velocity amplitudes. Here
we used the median level of the white noise in the range 3\,--\,5 mHz. The
accuracies for the BUSCA wavebands are 1.52\,mmag for '$UV_{\mbox{\footnotesize
B}}$', 1.53\,mmag for '$B_{\mbox{\footnotesize B}}$', 1.12\,mmag for 
'$R_{\mbox{\footnotesize B}}$' and 1.37\,mmag for '$NIR_{\mbox{\footnotesize 
B}}$', respectively.

\begin{figure}
\vspace{11cm}
\includegraphics{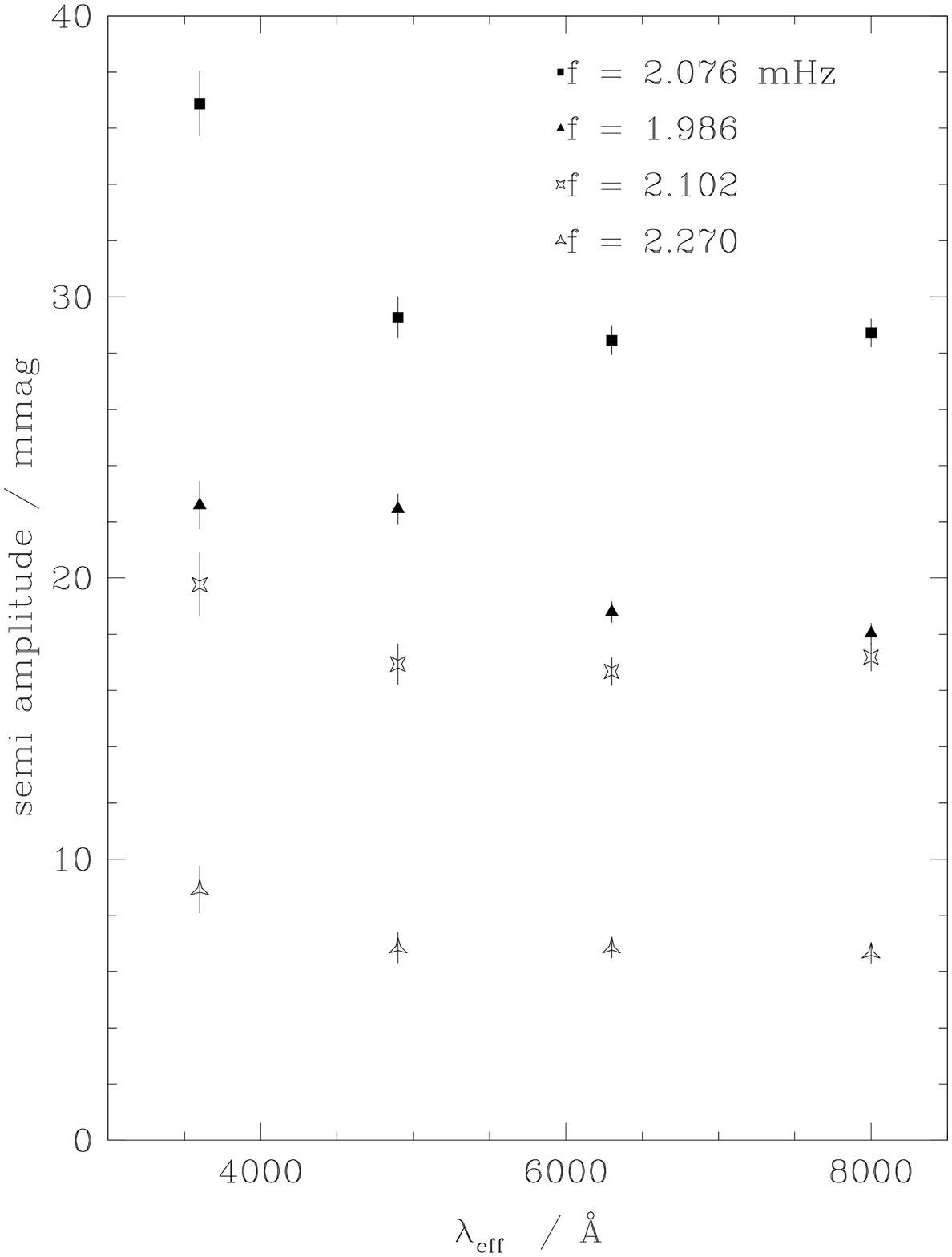}
\caption{Semi amplitudes of four frequencies as a function of effective 
wavelength. The error bars are 1$\sigma$ errors and are calculated by means of 
a $\chi^{2}$ method.}
\label{ampl}
\end{figure}

\begin{figure}
\vspace{11cm}
\includegraphics{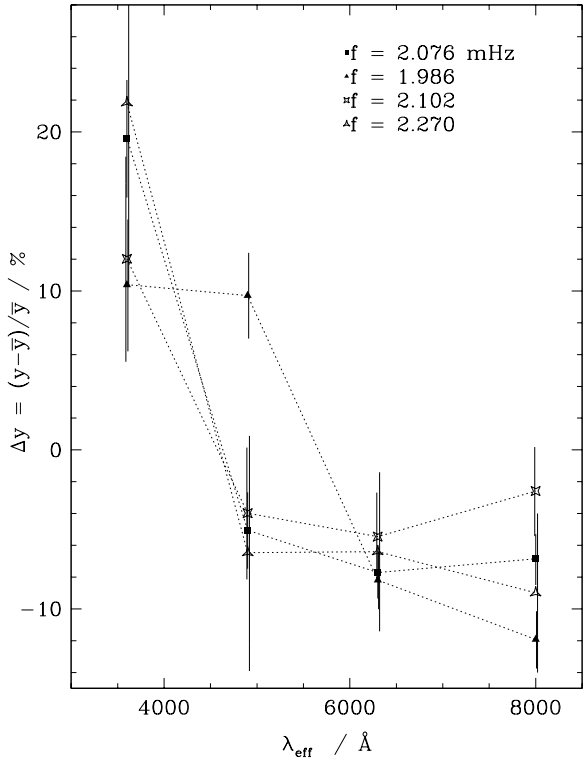}
\caption{Relative semi amplitudes of the four frequencies from Fig.~\ref{ampl}.
These are calculated by $\Delta\mathrm{y} = 
\left(\mathrm{y}-\mathrm{\bar{y}}\right)/\mathrm{\bar{y}}$. y is the semi 
amplitude of the wavebands. The error bars are the formal 1$\sigma$ errors of the
sine fit procedure.}
\label{relampl}
\end{figure}

Fig. \ref{ampl} shows the semi amplitudes of four selected frequencies as a 
function of effective wavelength of the bands. In Fig.~\ref{relampl}
we display the relative change of the semi amplitude of each waveband. The 
deviation with respect to the mean is largest for the '$UV_{\mbox{\footnotesize
B}}$' band. 
The other channels, considered separately, behave rather similar showing much 
smaller deviations from the mean brightness. This is explained 
through the fact that the '$UV_{\mbox{\footnotesize B}}$' band lies blueward to 
the Balmer jump and the other redward of it. The opacity changes a lot across 
this wavelength range and thus the stellar flux originates from different 
atmospheric depths.

Furthermore, we used the phases which are delivered by the sine fit 
procedure (see Table \ref{freqb} and \ref{freqc}) in order to test whether there 
is any wavelength dependency. The phase values are normalized to unity. 
Fig. \ref{phase} 
shows the deviations of the phases with respect to the mean value of all four 
wavebands for the four selected frequencies of Fig. \ref{ampl}. The error bars
are calculated from
\begin{equation}
  \mathrm{A} = \mathrm{Re}\{\mathrm{A}\}\,\mathrm{sin}\phi.\\
\end{equation}
The phase error $\Delta\phi$ can be calculated from the amplitude uncertainty 
$\Delta$A according to
\begin{equation}
  \Delta\phi =
  \mathrm{arctan}\left(\frac{\Delta\mathrm{A}}{\mathrm{A}}\right).\\
\end{equation}
The two strongest frequencies at 2.076 and 1.986\,mHz show no wavelength
dependent behavior within the error margins. The deviations for the other two
frequencies from the mean value are slightly larger than the formal errors.
However, we regard this as insignificant, and conclude that we cannot find 
any wavelength dependency of the phases of the modes.

\begin{figure}
\vspace{11cm}
\includegraphics{phase.eps}
\caption{Deviation of the phases from the mean value calculated from
the four frequency bands for the frequencies given in Fig.~\ref{ampl}. The error
bars are calculated from the amplitude errors as described in the text.}
\label{phase}
\end{figure}

\section{Comparison with previous investigations\label{comp}}

\begin{table*}
\caption[]{Radial velocities (RV) of PG\,1605$+$072 derived for 
$\mathrm{H}_{\beta}$ and $\mathrm{H}_{\gamma}$\label{radvel} and for comparison 
of the other radial velocity studies. Woolf, Jeffery \& Pollacco (\cite{wool02}) 
measured their radial velocities from shifts of the whole spectrum. The radial 
velocities of O'Toole et al. (\cite{otoo02}) are measured for all Balmer lines 
(velocity error: $\approx$ 0.4\,km/s).}
\begin{tabular}{ccc|ccc|cc|cc}
\hline
        & $\mathrm{H}_{\beta}$ &  &  & $\mathrm{H}_{\gamma}$ & & Woolf et al.
	(\cite{wool02}) & & O'Toole et al. (\cite{otoo02}) & \\
\hline
 f & P & RV & f & P & RV & P & RV & P & RV \\
 $\mathrm{[mHz]}$ & [s] & [km/s] & [mHz] & [s] & [km/s] & [s] & [km/s] & [s] &
 [km/s]  \\
\hline
 2.078 & 481.28 & 12.7 & 2.076 & 481.66 & 14.3 & 481.7/475.3 & 3.9/4.0 &
 480/475 & 4.3/8.5  \\
 2.756 & 362.89 &  8.0 & 2.753 & 363.21 &  6.5 & 363.2/366.2 & 3.0/6.1 & 365 & 
  7.2 \\
 1.985 & 503.79 &  7.9 & 1.978 & 505.77 &  7.2 & 502.0       & 3.9     & 504 & 
  4.1 \\
       &        &       &       &        &       & 527.1       & 2.7     &     &
            \\
\hline
\end{tabular}
\end{table*}

\begin{table*}
\caption[]{Frequencies and corresponding periods as well as amplitudes for
PG\,1605$+$072 in the first two BUSCA wavebands ('$UV_{\mbox{\footnotesize B}}$' 
and '$B_{\mbox{\footnotesize B}}$'). The errors of the periods and amplitudes
are the formal fit errors from the sine fit procedure. The left column shows 
the corresponding values derived in Kilkenny et al.
(\cite{kilk99}) as a comparison to the values found in this work. The two
frequencies marked with a star can not be resolved in our data but {\bf can} 
in the data
of Kilkenny (see Sect.~\ref{comp} for the discussion). The phases derived within 
the sine fit procedure are denoted by $\phi$ and are discussed in
Sect.~\ref{anal} as well as the determination of the phase errors.\label{freqb}}
\begin{tabular}{ccc|c|ccc|ccc}
\hline
Kilk.'99 & &                         &
        & '$UV_{\mbox{\footnotesize B}}$' &     &     & '$B_{\mbox{\footnotesize B}}$' &    &      \\
\hline
$\mathrm{f_{Kilk99}}$ & $\mathrm{P_{Kilk99}}$ & $\mathrm{A_{Kilk99}}$ & f & P & 
 A & $\phi$ & P & A & $\phi$ \\
 $\mathrm{[mHz]}$ & [s] & [mmag] & [mHz] & [s] & [mmag] &  & [s] & [mmag] &  \\
\hline
 2.0758 & 481.75 & 27.4 & 2.0760 & 481.69$\pm$0.02 & 36.88$\pm$0.12 & 
 0.684$\pm$0.006 & 481.69$\pm$0.02 & 29.27$\pm$0.07 & 0.700$\pm$0.008 \\
 1.9853 & 503.70 &  3.3 & 1.9861 & 503.51$\pm$0.03 & 22.59$\pm$0.09 & 
 0.953$\pm$0.010 & 503.60$\pm$0.02 & 22.45$\pm$0.06 & 0.951$\pm$0.010 \\
 $\star$2.1017 & 475.82 & 15.4 & 2.1020 & 475.74$\pm$0.04 & 19.76$\pm$0.11 & 
 0.026$\pm$0.011 & 475.82$\pm$0.03 & 16.94$\pm$0.07 & 0.000$\pm$0.013 \\
 $\star$2.1033 & 475.45 & 15.9 &        &                   &                 &
         &                   &             &     \\
 2.7613 & 362.15 &  1.8 & 2.7631 & 361.92$\pm$0.03 & 16.01$\pm$0.16 & 
 0.051$\pm$0.014 & 361.97$\pm$0.03 & 10.54$\pm$0.07 & 0.043$\pm$0.021 \\
 2.7663 & 361.49 &  2.0 & 2.7668 & 361.43$\pm$0.06 & 10.94$\pm$0.15 & 
 0.192$\pm$0.020 &                   &             &     \\
 --      & --     & --   & 2.7554 &                   &                 & 
         & 363.10$\pm$0.02 & 14.97$\pm$0.07 & 0.080$\pm$0.015 \\
 2.2701 & 440.51 &  5.2 & 2.2700 & 440.52$\pm$0.05 &  8.92$\pm$0.08 & 
 0.144$\pm$0.025 & 440.54$\pm$0.04 &  6.85$\pm$0.05 & 0.148$\pm$0.032 \\
 1.8914 & 528.70 & 13.9 & 1.8915 & 528.69$\pm$0.09 &  7.65$\pm$0.09 & 
 0.675$\pm$0.029 & 528.79$\pm$0.05 &  8.25$\pm$0.05 & 0.643$\pm$0.027 \\
 2.7173 & 368.01 &  0.6 & 2.7191 & 367.77$\pm$0.06 &  7.45$\pm$0.10 & 
 0.897$\pm$0.029 &                   &             &     \\
 2.3920 & 418.05 &  2.2 & 2.3921 &                   &                 &
         & 418.05$\pm$0.04 &  7.87$\pm$0.05 & 0.833$\pm$0.028 \\
 --      & --     & --   & 4.0748 &                   &                 &
         & 245.41$\pm$0.02 &  4.39$\pm$0.05 & 0.549$\pm$0.049 \\
\hline
\end{tabular}
\end{table*}

\begin{table*}
\caption[]{Frequencies and corresponding periods as well as amplitudes for
PG\,1605$+$072 in the last two BUSCA wavebands ('$R_{\mbox{\footnotesize B}}$' 
and '$NIR_{\mbox{\footnotesize B}}$'). The errors of the periods and amplitudes
are the formal fit errors from the sine fit procedure. The left column shows 
the corresponding values derived in Kilkenny et 
al. (\cite{kilk99}) as a comparison to the values found in this work. The phases 
derived within the sine fit procedure are denoted by $\phi$ and are discussed 
in Sect.~\ref{anal} as well as the determination of the phase 
errors.\label{freqc}}
\begin{tabular}{ccc|c|ccc|ccc}
\hline
Kilk.'99 &  &                        &
        & '$R_{\mbox{\footnotesize B}}$' &    &     & '$NIR_{\mbox{\footnotesize B}}$' &   &    \\
\hline
$\mathrm{f_{Kilk99}}$ & $\mathrm{P_{Kilk99}}$ & $\mathrm{A_{Kilk99}}$ & f & P & 
A & $\phi$ & P & A & $\phi$ \\
 $\mathrm{[mHz]}$ & [s] & [mmag] & [mHz] & [s] & [mmag] &  & [s] & [mmag] &  \\
\hline
 2.0758 & 481.75 & 27.4 & 2.0759 & 481.71$\pm$0.01 & 28.45$\pm$0.05 &
 0.691$\pm$0.006 & 481.73$\pm$0.02 & 28.72$\pm$0.07 & 0.684$\pm$0.007 \\
 1.9853 & 503.70 &  3.3 & 1.9858 & 503.58$\pm$0.01 & 18.79$\pm$0.04 & 
 0.948$\pm$0.009 & 503.61$\pm$0.02 & 18.02$\pm$0.05 & 0.936$\pm$0.011 \\
 2.1017 & 475.82 & 15.4 & 2.1020 & 475.76$\pm$0.02 & 16.68$\pm$0.05 &
 0.028$\pm$0.010 & 475.70$\pm$0.03 & 17.18$\pm$0.07 & 0.057$\pm$0.012  \\
 --      & --     & -- & 2.7530 & 363.23$\pm$0.01 & 17.15$\pm$0.04 &
 0.054$\pm$0.010 &                  &       &                  \\
 --      & --     & -- & 2.7866 & 358.87$\pm$0.03 &  6.22$\pm$0.04 &
 0.213$\pm$0.026 &                  &       &                  \\
 2.7427 & 364.60 & 15.1 & 2.7427    &               &               &
         & 364.60$\pm$0.02 & 10.37$\pm$0.06 &  0.043$\pm$0.019  \\
 --      & --     & -- &  2.7637    &                 &               &
         & 361.84$\pm$0.02 & 13.40$\pm$0.06 &  0.057$\pm$0.015  \\
 2.2701 & 440.51 &  5.2 & 2.2700 & 440.56$\pm$0.03 &  6.85$\pm$0.04 &
 0.133$\pm$0.024 & 440.60$\pm$0.04 &  6.66$\pm$0.05 & 0.183$\pm$0.030 \\
 1.8914 & 528.70 & 13.9 & 1.8914 & 528.71$\pm$0.04 &  6.79$\pm$0.04 &
 0.670$\pm$0.024 & 528.90$\pm$0.06 &  6.88$\pm$0.05 & 0.625$\pm$0.029 \\
 2.3920 & 418.05 &  2.2 & 2.3920 & 418.06$\pm$0.03 &  5.80$\pm$0.04 &
 0.811$\pm$0.028 & 417.94$\pm$0.04 &  6.44$\pm$0.05 & 0.821$\pm$0.031 \\
 4.0618 & 246.19 &  1.2 & 4.0624 & 246.16$\pm$0.01 & 4.47$\pm$0.04 &
 0.458$\pm$0.036 & 246.17$\pm$0.02 &  5.38$\pm$0.05 & 0.413$\pm$0.037 \\
\hline
\end{tabular}
\end{table*}

Other groups have carried out similar studies already, either time-series 
spectroscopy (O'Toole et al. \cite{otoo00}, O'Toole et al. \cite{otoo02}, Woolf 
et al. \cite{wool02}) or photometry (Kilkenny et al. \cite{kilk99}). O'Toole et 
al. (\cite{otoo00}) have collected monochromatic photometric data, too, but they
are not strictly simultaneous. We shall compare our results with their results 
in this section.

The photometric multi-site campaign of Kilkenny et al. (\cite{kilk99}) 
discovered more than 50 frequencies in their
frequency analysis of PG\,1605$+$072. The five frequencies (2.076, 1.9865,
2.102, 2.2695 and 1.8912\,mHz) detected in all BUSCA bands of our data were 
already present in their data. Some of the others which we found additionally 
(e.g. 4.0631\,mHz) can be identified with frequencies from Kilkenny et al. 
(\cite{kilk99}) and others can not (in particular at 2.7866 mHz).
{\bf A closer inspection of these frequencies shows that all of them are very
close to one-day aliases ($\pm$\,11.57\,$\mu\mathrm{Hz}$) of the extracted
frequencies of our or of Kilkenny's data to within the frequency resolution of 
5.68\,$\mu\mathrm{Hz}$.
} 
%It is 
%difficult to identify frequencies reliably in the crowded frequency range 
%2.74\,--\,2.78\,mHz and, therefore, 
Small deviations of our measurements from those of Kilkenny et al. 
(\cite{kilk99}) can be explained by the shorter time spanned by our data 
($\approx$ 48\,h compared to $\approx$ 217.6\,h, frequency resolution: 
5.68\,$\mu\mathrm{Hz}$ compared to 1.28\,$\mu\mathrm{Hz}$).
In contrast to our analysis, Kilkenny et al. (\cite{kilk99}) were able to 
resolve closely spaced frequencies, e.g. at 2.1017\,$\mu$Hz and 
2.1033\,$\mu$Hz 
(see Table \ref{freqb}). Since 1997 the relative power within one blend of
frequencies may have changed so that the frequency of unresolved peaks in the 
periodogram changes slightly because the contribution of the weaker feature at 
that time may have grown in the meantime. We conclude that the frequencies
measured in the BUSCA light curves are consistent with those of Kilkenny et al. 
(\cite{kilk99}). The frequencies are stable in time to within our measurement
errors.

Also the frequencies discovered in the radial velocity curves are consistent 
with both the BUSCA and the Kilkenny values. The time basis of the 
spectroscopic data is too short in comparison to the long-term photometry to 
resolve the frequencies. This is also true for most of the other spectroscopic 
feasibility studies mentioned before. Only O'Toole et al. (\cite{otoo02}) had a
sufficiently long baseline of $\approx$ 70\,d. Thus, we discuss only the 
relative distribution of the power of the pulsation frequencies. As can be
seen in Fig. \ref{LSpHb} the main power is located around 2.076\,mHz. 
Significant power arises in the frequency range 2.74\,--\,2.78\,mHz but it falls 
off compared to the former.
The same distribution of pulsation power was detected by the multi-site
photometric campaign of Kilkenny et al. (\cite{kilk99}) who carried out their
observations in April and May 1997. Two years later, in July and August 1999,
O'Toole et al. (\cite{otoo00}) recovered the same frequency pattern but 
discovered that the amplitudes had changed {\bf and, in particular, that} no 
power was detected at 2.076\,mHz. Another year later (in May 2000) O'Toole et 
al. (\cite{otoo02}) and 
Woolf, Jeffery \& Pollacco (\cite{wool02}) gathered data again. Both groups 
discovered a drastic change in the relative power distribution: The power 
at 2.076\,mHz appeared again and was at that time weaker than that in the 
frequency range 2.74\,--\,2.78\,mHz. This means that the power switches within a 
few years. Our data which were taken in May 2001, i.~e. another year later, 
show that the power distribution switched back to that of 1997 measured by 
Kilkenny et al. (\cite{kilk99}).

\section{Conclusion\label{conc}}

Our spectroscopic feasibility study based on data taken in May 2001 confirms 
earlier studies from observations in July/August 1999 and May 2000. All these
investigations have shown that it is possible to measure accurate radial 
velocity curves for PG\,1605$+$072. Three to five frequencies could be detected 
in the periodograms. The detected frequencies found in the radial velocity
studies are very stable (see Table \ref{freqb} and \ref{freqc}). However, the
amplitudes of the frequencies show drastic temporal changes over a few years. We
also have simultaneously measured photometric variations of PG\,1605$+$072 in 
four wavebands for the first time. The amplitudes in the 
'$UV_{\mbox{\footnotesize B}}$' band are found to be significantly larger than 
at larger wavelengths ('$B_{\mbox{\footnotesize B}}$', '$R_{\mbox{\footnotesize 
B}}$' and '$NIR_{\mbox{\footnotesize B}}$'). No phase shifts between the four 
light curves can be detected.

The time basis of all observations is much too short to resolve the full 
frequency spectrum of this star. More than 50 frequencies have been resolved in
the long photometric study of Kilkenny et al. (\cite{kilk99}). Consequently, a 
new data set with a longer time basis is needed. For 
that reason we organised a combined photometric and spectroscopic multi-site 
campaign (PI: U. Heber), termed the MSST ({\bf M}ulti {\bf S}ite 
{\bf S}pectroscopic {\bf T}elescope) project (Heber et al. \cite{hebe02}). 
These observations covered a period of one month in May/June 2002. This project 
is unique in that it combines photometric and spectroscopic data. The data 
still have to be reduced and analysed. A lot of work remains to be 
done. Hopefully, we will be able to identify pulsation modes within this large 
data set and carry out a detailed asteroseismological investigation of 
PG\,1605$+$072. The feasibility study presented in this paper in concert with
all other equivalent studies mentioned in Section \ref{comp} have paved the way
for this aim.

\begin{acknowledgements}
The authors would like to thank the Calar Alto staff for their support during 
our observation run in May 2001. Financial support by the \emph{Deutsches 
Zentrum f\"ur Luft- und Raumfahrt} (\emph{DLR}) under grant 50\,OR\,96029-ZA is 
gratefully acknowledged. We also thank the \emph{Deutsche Forschungs 
Gemeinschaft} (\emph{DFG}) for a travel grant (HE\,1356/33-1) to Calar Alto 
observatory. BUSCA was realised with financial support by the german 
\emph{Bundesministerium f\"ur Bildung, Wissenschaft, Forschung und Technologie} 
(\emph{BMBF}) through grant 0\,3BN114\,(4) of Verbundforschung 
Astronomie/Astrophysik. We thank Simon O'Toole for his valuable comments on the 
text and contributing a routine to calculate the errors of the amplitudes and 
phases.
\end{acknowledgements}

\clearpage

\end{document}